\documentclass[12pt]{article}
\usepackage{mathtext}
\usepackage{graphicx}
\usepackage[english]{babel}
\usepackage[utf8]{inputenc}
\usepackage{amsfonts}
\usepackage{amssymb}
\usepackage{amsmath}
\usepackage{float}
\newcommand{\e}{\epsilon}

\begin{document}
\title{Reliability of optical gate array for C-Sign operator}
\author{M.Lemus\thanks{RUDN, marianojlemush@gmail.com}, Y.Ozhigov\thanks{MSU, VMK, ozhigov@cs.msu.su}, N.Skovoroda\thanks{MSU, VMK, chalkerx@gmail.com}}
\maketitle
 03.65,  87.10\\

\begin{abstract}
We study optical gate array of KLM model of quantum computer for C-Sign operator, which contains linear elements and two nonlinear phase shifts. Linear elements and photon counting are taken to be ideal, whereas nonlinear phase shifts are subject of realistic and unavoidable factors of decoherence: the immediate photon leak from cavities and mediated leakage with the atom, which flies out of the cavity and plays the role of ancilla. We use JC model with Lindblad master equation to estimate the dependence of the validity of C-Sign operator from the rate of photon leak and estimate the limitation, which this validity imposes to quantum computations.
\end{abstract}

\section{Introduction and background}

	Quantum mechanics gives us the new paradigm of computations: the peculiar type of information processing realized by quantum computers, called quantum computations. It has been shown by \cite{SH} and \cite{G} that for some specific tasks, quantum computation allows for faster performance than it is theoretically possible in a classical computer. Quantum information and encryption provides a secure way of transmitting information based on the wavefunction collapse property of quantum systems and the use of quantum encryption algorithms.  One of the main challenges of the field is the physical realization of a quantum computer. One of the proposals, the KLM scheme \cite{KLM}\cite{KM}\cite{ML}, based on linear optical components, has paved the way for different implementations based on quantum optics. 
	
	The photon, being a chargeless and massless particle, is a very adequate candidate as the carrier of quantum information. The advantages of using photons include an easy and low-loss way to transport them using optical fiber, their relatively well understood quantum properties, and the existence of well-developed techniques for generating and measuring photons.

	Linear quantum optics has proven to be very good playground for experimental and theoretical developments in the realization of quantum computers. The photons allow for at least two viable representations of quantum information in form of the dual rail and the polarization representation. The parametric down conversion \cite{MP} and the single atom microlaser \cite{H} allow us to produce single photons on demand, ready for computation. In addition to that, the KLM scheme provides a mechanism from which all the single qubit gates can be constructed using only mirrors, phase shifters and beamsplitters, which are all common and well understood. Even the nonlinear effects needed to build the CNOT gate can be obtained by the non-deterministic nonlinear shift (NS) gate. The associated decoherence of those components is low, meaning that most of the information loss comes from the photon storage between operations and from the photon measurement. Both of these can be reduced using optical resonators to store light in the infrared and microwave domain, and using solid state photoelectric detectors based in semiconductors to measure the qubits.

	The main problem with the KLM scheme is scalability; this means that the resources needed for computation should increase linearly with the number of qubits used.  The fact that the NS gate has a low success probability makes successive applications of it unreliable, as the probability of success quickly decreases exponentially. This effect can be countered using quantum teleportation gates as described in \cite{GC}, but the amount of entangled pairs of qubits needed to make the CNOT gate work with 95\% reliability is in the order of 100 pairs per gate \cite{R} which limits the scheme scalability. 

	Another solution to the scalability problem is to extend the scheme to include nonlinear components. One way of realizing the NS gate is using nonlinear Kerr media to produce cross phase modulation, theoretically solving the main problem of the optical quantum computer model. Nevertheless, there is always some absorption associated with this nonlinearity, and it has been estimated \cite{NC} that in the best known case, approximately 50 photons must be absorbed to get one to experience a $\pi$ cross phase modulation. This means that currently, the use of Kerr media isn’t a solution for the scalability problem.

	Cavity QED poses itself as a solution to the challenge of scalable qubit entanglement, both for qubits represented as atom states with interaction mediated by photons as shown by \cite{RK}, as well as for qubits represented by photons interacting through atoms inside a cavity. A single atom in a cavity is enough to produce an entangled state between photons, as shown experimentally by \cite{T}. Recently, a setup by \cite{A} realizes the NS gate with an upper bound error probability of approximately 2.47\%. This work continues by generalizing it to explore how the inclusion of decoupling and dissipative effects change the potential reliability of the NS and C-Sign gates.

\section{Description of the physical system}

\begin{figure}
\begin{center}
\includegraphics[width=0.5\textwidth]{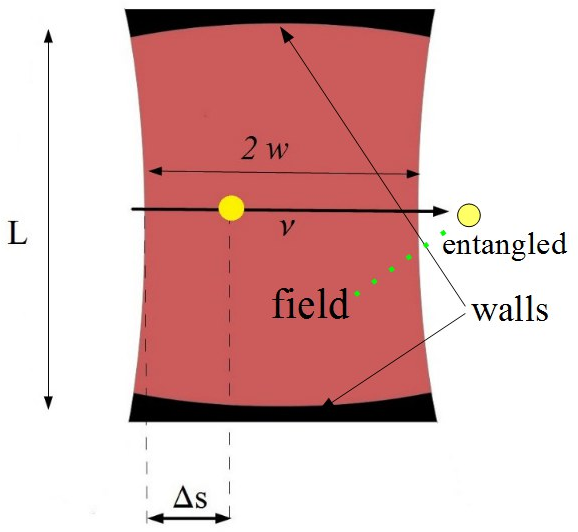}
\caption{NPS: optical cavity, which the two-level atom flies through. The energy of the field in the cavity does not exceed $2\omega_c$. We choose the appropriate time $\tau_0=\Delta s/v$ for finding the atom in the cavity for the realization of NFS.}
\end{center}
\end{figure}

Consider an optical Fabry-Perot cavity (as shown in Figure 1) with a two-level Rydberg atom flying through it with speed $v$. The geometrical parameter $w$ denotes the half-width of the beam waist at the center of the cavity and $L$ denotes the distance between the mirrors. Let the low and high energy states of the atom be denoted by $|g\rangle$ and $|e\rangle$ respectively an the $n$ photon state inside the cavity be $|n\rangle$. Suppose that the atoms inside the cavity are prepared initially in the $|g\rangle$ state, and the system is built so no more than two photons can enter the cavity. The Hilbert space associated with the system is $Span\{|g,0\rangle,|g,1\rangle,|g,2\rangle,|e,0\rangle,|e,1\rangle\}$. The evolution of such system is described by the Jaynes Cummings Hamiltonian (\cite{JC}):

\begin{equation}
H_{JC}=\omega_ca^+a+\omega_a\sigma_z/2+
\gamma(\sigma^+a+\sigma a^+)
\label{JC}
\end{equation}

where $a_i^+,a_i$ are operators of creation and annihilation of photons; $\sigma^+, \sigma$ are operators of excitation and relaxation of the atom in the cavity, and $\gamma$ is the coupling constant for interaction between the atom and the field ($\hbar $ is taken to be $1$). This Hamiltonian leaves invariant subspaces of the form $\Lambda_n = Span\{|g,n+1\rangle,|e,n\rangle\}$ and the space $\Lambda_g = Span\{|g,0\rangle\}$. By taking the detuning $\Delta = \omega_c-\omega_a$, $H_{JC}$, limited on  $\Lambda_n$ can be written as a $2\times2$ matrix 

\begin{equation}
H_{JC}=(n+1/2)\omega_c\begin{pmatrix}
  1 & 0\\
  0 & 1 \\
 \end{pmatrix}+
\begin{pmatrix}
  \Delta/2 & \sqrt{n+1}\gamma\\
  \sqrt{n+1}\gamma & -\Delta/2 \\
 \end{pmatrix}
\label{JCMatrix}
\end{equation}

The evolution of the system can be fully obtained by solving the eigenvalue problem for $H_{JC}$, which gives the eigenvalues for the energy:

\begin{equation}
\e_\pm=(n+1/2)\omega_c\pm\Omega_n/2
\label{EigenV}
\end{equation}

where $\Omega_n=\sqrt{\Delta^2+4\gamma^2(n+1)}$ is the generalized Rabi frequency of the system. The associated eigenstates are the dressed states of the system

\begin{equation}
\begin{array}{lcl} |+,n\rangle=cos(\theta_n)|g,n+1\rangle+sin(\theta_n)|e,n\rangle
 \\ |-,n\rangle=-sin(\theta_n)|g,n+1\rangle+cos(\theta_n)|e,n\rangle
  \end{array}
\label{EigenS}
\end{equation}

where $tan(\theta_n)=(2\sqrt{n+1}\gamma)/(\Omega_n-\Delta)$. For the system ground state $|g,0\rangle$, the energy eigenvalue is given by

\begin{equation}
\e_g=(-1/2)\omega_c-\Delta/2
\label{EigenVG}
\end{equation}

Considering the system as a part of an array of components, in which the number of photons is fixed, we turn our attention to only the interaction and free atom part of the Hamiltonian. It is useful then, to consider the interaction picture associated with $H_0=(a^+a+1/2)\omega_c$. In such picture, the evolution of the state according to (\ref{JC}) is

\begin{equation}
\psi(t)=\alpha_0e^{-it\e_g}|g,0\rangle+\sum_{j}(\alpha_{+,j}e^{-it\e_{+,j}}|+,j\rangle+\alpha_{-,j}e^{-it\e_{-,j}}|-,j\rangle)
\label{JCSolution}
\end{equation}

Let the system be initially prepared in a state of the form $\psi(0)=\alpha_0|g,0\rangle+\alpha_1|g,1\rangle+\alpha_2|g,2\rangle$. Then, equation (\ref{JCSolution}) shows that after a time $t$ the system evolves to the state

\begin{equation}
\begin{split}
\psi(t)=&\alpha_0e^{it\Delta/2}|g,0\rangle+ \\
	    &+\alpha_1[(cos(\Omega_0t/2)-icos(2\theta_0)sin(\Omega_0t/2))|g,1\rangle- \\
		&-isin(2\theta_0)sin(\Omega_0t/2)]|e,0\rangle+ \\
		&+\alpha_2[(cos(\Omega_1t/2)-icos(2\theta_1)sin(\Omega_1t/2))|g,2\rangle- \\
		&-isin(2\theta_1)sin(\Omega_1t/2)]|e,1\rangle
\end{split}
\label{JCSolution2}
\end{equation}

The dissipation inside the cavity can be handled using a master equation in the Lindblad form on the system's density matrix $\rho$

\begin{equation}
ih\dot{\rho}=[H,\rho ]+i\sum\limits_j\kappa_j(L_j\rho L_j^+-\frac{1}{2}\{ L_j^+L_j,\rho \})
\label{master}
\end{equation}

The main source of decoherence inside an optical cavity is the photon leakage through the mirrors, the effect of which can be described by the photon annihilation operator acting on the electric field state $L_1=a$ and the photon escape rate constant $\kappa_{ph}$, which depends on the $Q$ factor of the cavity and the wavelength of the photons. 
The resulting Lindblad equation is

\begin{equation}
ih\dot{\rho}=[H_{JC},\rho ]+i\kappa_{ph}(a\rho a^+-\frac{1}{2}\{ a^+a,\rho \}).
\label{JCmaster}
\end{equation}

\section{Mathematical model of C-Sign gate array}
\label{analythical}

\begin{figure}
\includegraphics[scale=0.5, viewport =-53 40 30 120]{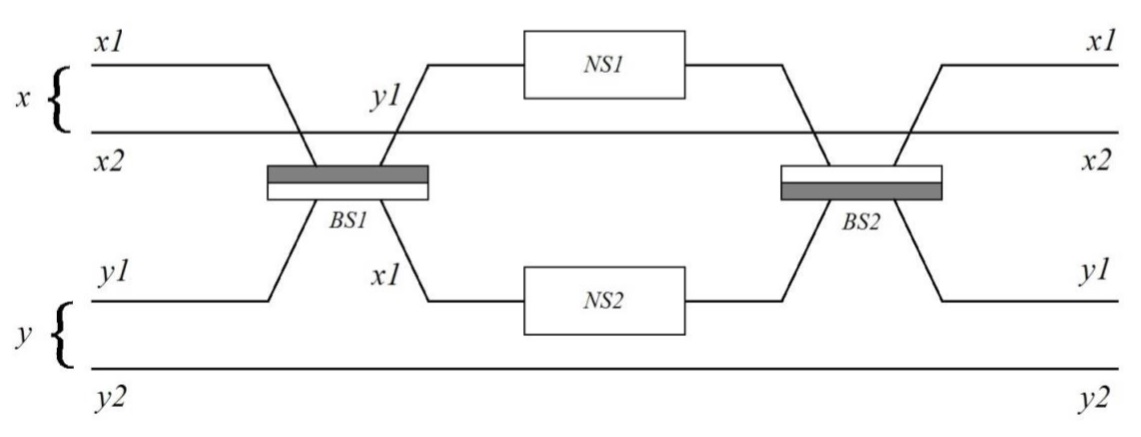}
\caption{C-Sign gate array}
\end{figure}

The C-Sign gate array, proposed in \cite{A} is shown in Figure 2. The wires are denoted by $x_1,x_2,y_1,y_2$ and each wire can contain $0,1$ or $2$ photons of the fixed frequency $\omega_c$ and circular polarization, with a single limitation: the number of photons that are present in the gate array simultaneously does not exceed $2$. The number of photons in any wire we denote by the same letter.

Decoherence comes from the non ideal work of photo detectors, for example, dark countings, photon leakage from the walls of the cavities and from the harmful entanglement between photons and atom, which can remain when the atom flies out of the cavity. In this work we ignore the errors of photo detectors and analyze only the photon leakage and harmful entanglement. We also assume that the linear optical elements work ideally, thus the single source of errors is connected with the work of optical cavities. 

The logical states of our computations are the spatial photon states in the dual rail representation. The qubits are denoted by $x$ and $y$ in Figure 2 and the computational basis is defined as

\begin{equation}
|0\rangle=|0\rangle_{1}|1\rangle_{2} \hspace{35pt}
|1\rangle=|1\rangle_{1}|0\rangle_{2}
\label{DualRail}
\end{equation}

The atomic states play the role of ancilla, which is prepared in the state $|g\rangle$ and in the ideal case must leave the gate array in the same state. Unfortunately, the ideal situation is unattainable in the given model even theoretically. We can only get an increasingly accurate approximation to it by choosing larger times of interaction between the field and the atom in an optical cavity. However, this increase of time will inevitably lead also to the increase in the probability of photon leakage out of the cavity. Our task is to find the optimal time for finding the atom in an optical cavity, depending on the intensity of photons leak.

The array consists in four wires, two idle ones $x_2,y_2$ and two wires  $x_1,y_1$ which are acted upon by the optical components. The idle wires cannot be disregarded, as the photons inside of them contribute to the overall energy of the system.

There are two linear gates: beam splitters BS1 and BS2, whose action in Fock space looks as

\begin{equation}
\begin{split}
\|n\rangle_{a_1}|m\rangle_{a_1}=\frac{1}{\sqrt{n!m!}}(a_1^+)^n(a_2^+)^m|0\rangle_{a_1}|0\rangle_{a_2}\longrightarrow \\
\longrightarrow\frac{1}{\sqrt{n!m!}}[\frac{1}{\sqrt{2}}(a_1^++a_2^+)]^n[\frac{1}{\sqrt{2}}(a_1^+-a_2^+)]^m|0\rangle_{a_1}|0\rangle_{a_2}
\end{split}
\label{BSplitter}
\end{equation}

and two optical cavities, which perform the nonlinear shifts: NS1, NS2. The nonlinear shift in the ideal case acts on Fock states of photons as

\begin{equation}
|0\rangle\rightarrow |0\rangle, \ |1\rangle\rightarrow |1\rangle,\ |2\rangle\rightarrow -|2\rangle
\label{NS}
\end{equation}
 
 This is realized by Jaynes-Cummins Hamiltonian (\ref{JC}).
The gates work sequentially, from the left to the right, as shown in the Figure 2. We assume that the beam splitters work instantly, whereas the NS gates work simultaneously, and their work lasts for the time $\Delta t$. Nonlinear shifts are realized by optical cavities with resonant frequency $\omega_c$, which is close to the frequency of atom transition from the excited to the ground state $\omega_a$ with a small detuning $\Delta$. The atom, initially prepared in the ground state, flies through the cavity along the path, and during the flight interacts with field inside the cavity via  (\ref{JCSolution2}). We interrupt this interaction by the deflation of photons from the cavity in the moment $t=\tau$ that is chosen in advance; the choice of $\tau$ determines the accuracy of the gate array. 

At this stage we meet the unavoidable decoherence factor, which comes from the residual entanglement between atom and field, mentioned above.

To choose $\tau$ we consider three arithmetic progressions
\begin{equation}
\begin{array}{lll}
&A: \ t_0^0,t_1^0,\ldots, &t_s^0,\ldots,\\
&B: \ t_0^1,t_1^1,\ldots, &t_n^1,\ldots,\\
&C: \ t_0^2,t_1^2,\ldots, &t_m^2,\ldots,
\end{array}
\label{progressions}
\end{equation}
where $A,B$ and $C$ consist of the sequential time moments, when the unitary evolution of JC transforms photonic states as $|0\rangle$ to $|0\rangle$, $|1\rangle$ to $|1\rangle$ and $|2\rangle$ to $-|2\rangle$ respectively. If we choose the time $\tau$, when these progressions approximately coincide, we will obtain the required duration.

The evolution of Fock states is shown in (\ref{JCSolution2}). Eigenvalues of the Hamiltonian blocks (\ref{JCMatrix}) for $n=0,1,2$ have the form
\begin{equation}
\begin{array}{lll}
&\e_g=(-1/2)\omega_c-\Delta/2\\
&\e_{\pm,0}=(1/2)\omega_c\pm\sqrt{\Delta^2+4\gamma^2}/2\\
&\e_{\pm,1}=(3/2)\omega_c\pm\sqrt{\Delta^2+8\gamma^2}/2
\end{array}
\end{equation}

Nonlinear dependence of amplitude from the photon number, called optical nonlinearity is a rare and exceptional phenomenon. A manifestation of this exclusivity is the fact that the match in progressions (\ref{progressions})  is only approximate. We could consider the generalization of operator (\ref{NS}): $
|0\rangle\rightarrow e^{ia}|0\rangle,\ |1\rangle \rightarrow e^{ib}|1\rangle, \ |2\rangle\rightarrow e^{ic}|2\rangle
$ , where $a+c\neq 2b$, and realize its particular case with only one separated optical cavity with flying atom by the choice of appropriate parameters $\Delta$ and the time $\tau$.

The nonlinearity means that for some $t>0$, integers $k_g,k_{\pm,0}, k_{\pm,1}$ and $a+c\neq 2b$ (modulo $2\pi$) (nonlinearity of the phase on photon number) the following four equations are true:
$
\e_gt=2\pi k_g +a, \ 
\e_{\pm,0}t=2\pi k_{\pm,0} +b,\ \e_{\pm,1}t=2\pi k_{\pm,1} +c.
$

By inspection of equation (\ref{JCSolution2}), it is clear that $e^{ib}$ and $e^{ic}$ must be real numbers, which means that $b,c\in\{0,\pi\}$ (modulo $2\pi$). We note that for the zero detuning ($\Delta=0$) the impossibility of the exact coincidence between elements of the sequence $B$ and $C$ is evident, because $\sqrt{2}$ is not rational and thus the differences of these progressions are incommensurable. We can only choose the value of $\tau$ that is maximally close to the both progressions. The more accuracy we wish to obtain the larger $\tau$ will be. We are only interested in the accuracy of the scheme, not the time of its execution. The value  $\tau$ can thus only affect the quality of the circuit via the second factor of decoherence - the leak of photons through the walls of cavities.

For the detuned case, the nonlinearity condition can be satisfied if we take such $d=\Delta/\gamma$ so that $\sqrt{4+d^2}/\sqrt{8+d^2}$ is rational; that can be guaranteed if we find the rational solutions of the equation $Y^2-X^2=1$, for which there is dense set of appropriate values for the parameter $d$. To achieve nonlinearity it is then sufficient that $a\neq c$ (modulo $2\pi$). That we can achieved by almost any arbitrary choice of parameters among these solutions. For the C-Sign case we need more, (\ref{NS}) is realized only when $a=c+\pi$ (modulo $2\pi$). The numerical computations with Mathematica shows that there is no solution of such a system. The best we can do is to find the approximate solution.

The evolution with the leak of photons gives us the mixed state even if the initial state was pure; we thus must use the density matrix $\rho$ for the description of the evolution. We use master equation

The evolution of the state in the work of our gate array thus can be described by the equation (\ref{JCmaster}), where the Hamiltonian $H$ and $\kappa_{ph},\kappa_{at}$ are piecewise constants; namely, $H$ coincide with the Hamiltonian, corresponding to the action of linear elements (for the simplicity, we can apply the corresponding transformation (\ref{BSplitter}) directly in the corresponding time) without dissipation, and in the time frame when the field interacts with the atoms we set $H=H_{JC}$ and take the actual value of the $\kappa$ coefficients. 

We use the distance $\delta (\rho_0,\rho )=\|\rho_0-\rho\| $ between the ideal density matrix $\rho_0$ and the real density matrix $\rho$ at the end of computation as the criterion of the validity of the gate array. This value $\delta$ depends on the intensity $\kappa$ of photon leak, and we can find this dependence for the different choice of $\tau_0$.

\section{Numerical simulation of the density matrix dynamics}

All gates, except for the main part of the $NS$ gates, are considered being ideal (both beamsplitter gates and the phase shift inside the $NS$ gates, if enabled).

The $NS$ gates simulation was performed in the assumption that both of them have identical time. The simulation operates with 6 particles (4 rails and 2 atoms): $x_1$, $x_2$, $y_1$, $y_2$, $a1$, $a2$. There are $19$ total possible states in the system, because all the other states are unreachable when using a valid input. The simulation deals only with those $19$ reachable states.

\subsection{The Hamiltonian for the $NS$ gate}

The Hamiltonian has to take into an account all the particles in the system. $x2$ and $y2$ receive only the phase shift over time with an amplitude which is assumed to be equal to $\omega_c$. Detuning $\Delta$ is equal to $\omega_a - \omega_c$.

The total Hamiltonian for the main part of the $NS$ gate (excluding the phase shift) has the following form:
\begin{equation}
\begin{array}{ll}
H_{NS} = & \omega_c (x_1^+x_1^- + x_2^+x_2^- + y_1^+y_1^- + y_2^+y_2^-) + \omega_a (a_1^+a_1^- + a_2^+a_2^-) + \\ & g (a_1^+x_1^- + x_1^+a_1^- + a_2^+x_2^- + x_2^+a_2^-)
\end{array}
\label{H}
\end{equation}

Variable $g$ could be excluded from the parameters list, if all the variables are taken relative to $g$. When doing so, the main parameters for the calculation are the following: $w_a/g, \Delta/g$ (or $\Delta/w_a$), $t = T \sqrt{2} g / \pi$ (where $T$ is the absolute $NS$ gate duration). $h$ is taken equal to $1$.

Also one extra parameter is $phs$, that denotes enabling ($phs = 1$) or disabling ($phs = 0$) a phase shift gate at the end of both $NS$ gates.

From here on, it is taken $g = 0.1$ (does not affect anything as all parameters are relative to $g$), $wc = g*\frac{5.11}{3.41}*10^6$. Values for detuning $\Delta$ that are used below are from $0$ to $5g$, which is five orders of magnitude lower than the cavity frequency.

\subsection{Lindblad operators for photon leakage}

We simulate the photon leakage from the cavities using the Lindblad master equation. With the inclusion of the Lindblad master equation in the diagonal form, the simulation equation takes the following form:
\begin{equation}
\begin{array}{lll}
\rho_{t + \delta t} &=& U_{\delta t}^*\rho_{t}U_{\delta t}
 + \delta t \sum\limits_{i}(L_i \rho L_i^* - \frac{1}{2}(L_i^* L_i \rho + \rho L_i^* L_i)),\\
U_{\delta t} &=& e^{-\dfrac{i \cdot \delta t}{\hbar} H}.
\end{array}
\label{modeleq}
\end{equation}
Lindblad operators for the photon leakage inside the $NS$ gates have the following form:
\begin{equation}
\begin{aligned}
L_1 = l_y * x_1^- \\
L_2 = l_y * x_2^- \\
\end{aligned}
\label{jchlr}
\end{equation}
In the following equations $l_y$ is the photon leakage coefficient. For the simulation, we take it relative to the photon-atom interaction strength $g$, so this adds one more parameter to our model: $l_y/g$.

\subsection{Calculating the error rate}

Error rate $error$ takes values from $0$ to $1$ and is equal to $1$ minus the gate validity. The gate validity from here on is calculated as  $max(|eigenvalues(P_{expected}-P_{result})))$ where $P_{test} = \frac{1}{4}(|x_1\rangle + |y_1\rangle + |x_2\rangle + |y_2\rangle)(\langle x_1|+ \langle y_1| + \langle x_2| + \langle y_2|)$ density matrix was used as the input. Also other random valid $P_{rand}$ density matrices with the total number of photons equal to $2$ (gate restriction) were tested as input, and the result corresponded with the result for $P_{test}$.

\subsection{The case of zero detuning}

\begin{figure}[H]
\begin{center}
\includegraphics[width=1.00\textwidth]{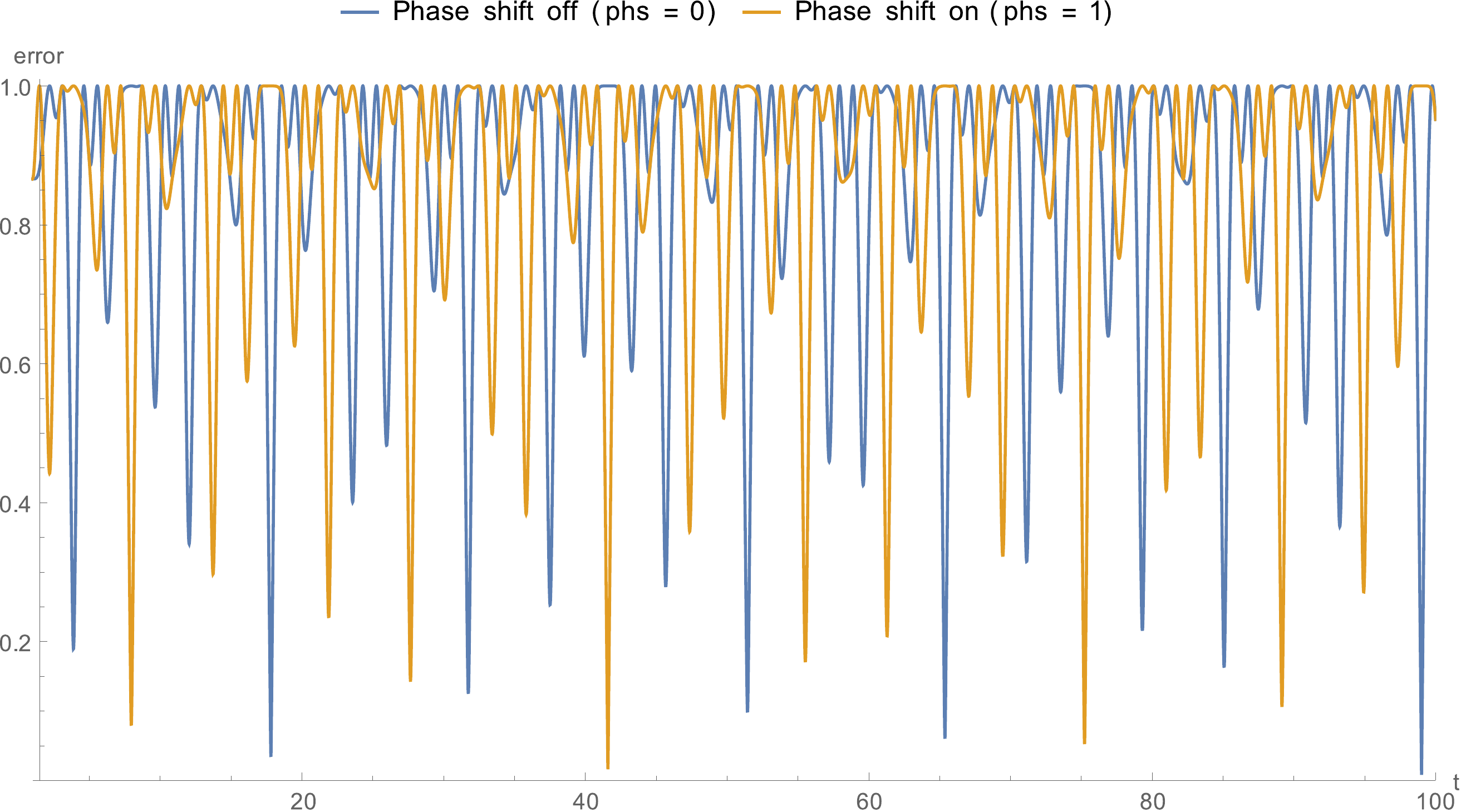}
\end{center}
 \caption{ \label{fig:errortime} Error over the selected NS gate time, with and without the phase shift, without detuning. Optimal values are $3$, $7$, $17$, $41$, $99$. Photon leakage is disabled.}
\end{figure}

In the case of zero detuning, the only viable parameter sets are those with whole values of $t$. That is explained in section \ref{analythical}, and could be also seen from the figure \ref{fig:errortime}.

\begin{figure}[H]
\begin{tabular}{c c}
\includegraphics[width=0.52\textwidth]{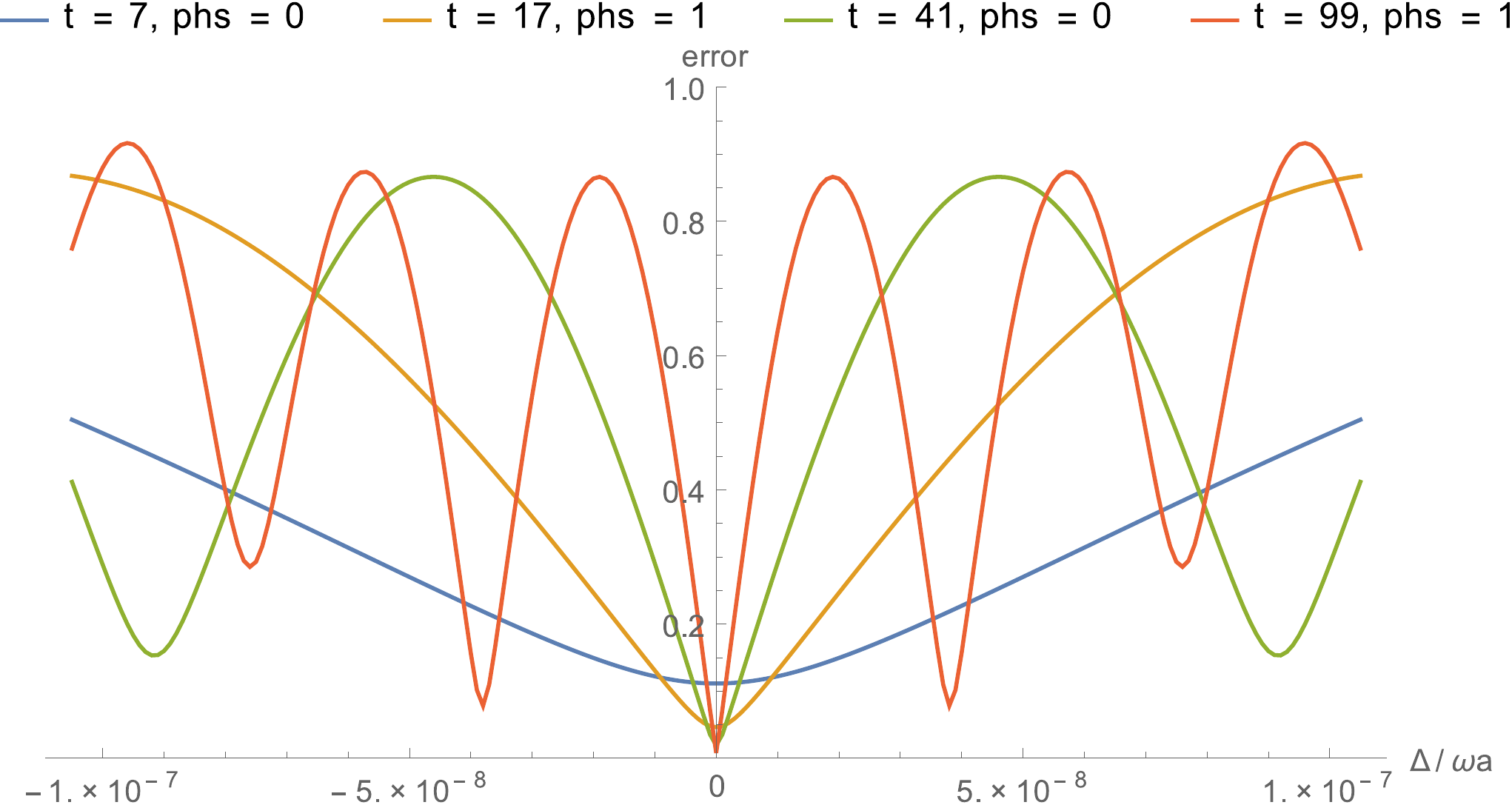} &
\includegraphics[width=0.48\textwidth]{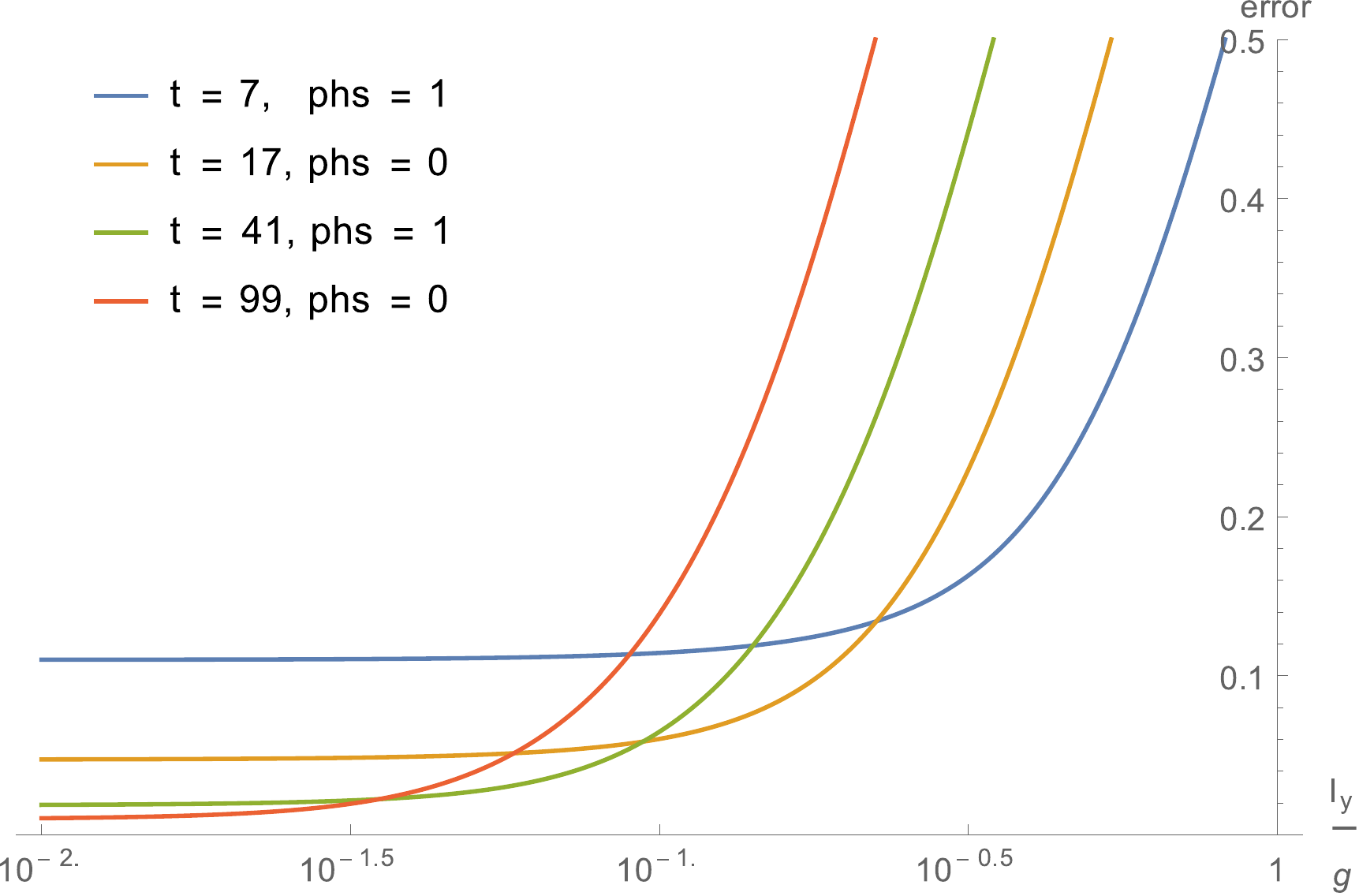} \\
Error over detuning & Error over photon leakage \\
& (logarithmic scale on leakage) \\
\end{tabular}
 \caption{ \label{fig:detuning-0} Error rates over detuning and photon leakage for the optimal cases. }
\end{figure}

The higher the $NS$ gate duration is, the more precise match we can find, and the lower is the error rate, but as could be seen on figure \ref{fig:detuning-0}, the faster does the error rate grow over increasing the photon leakage and/or the inaccurate detuning setting (which is the same as the detuning itself for the case when the desired detuning is equal to $0$). There are no reasons for choosing higher durations over lower ones, if the gate validity does not improve, so we treat as ``optimal'' only those sets of parameters that lower the error rate compared to the best found parameters set with lower time values.

\subsection{The case of non-zero desired detuning}

\begin{figure}[H]
\begin{center}
\includegraphics[width=0.7\textwidth]{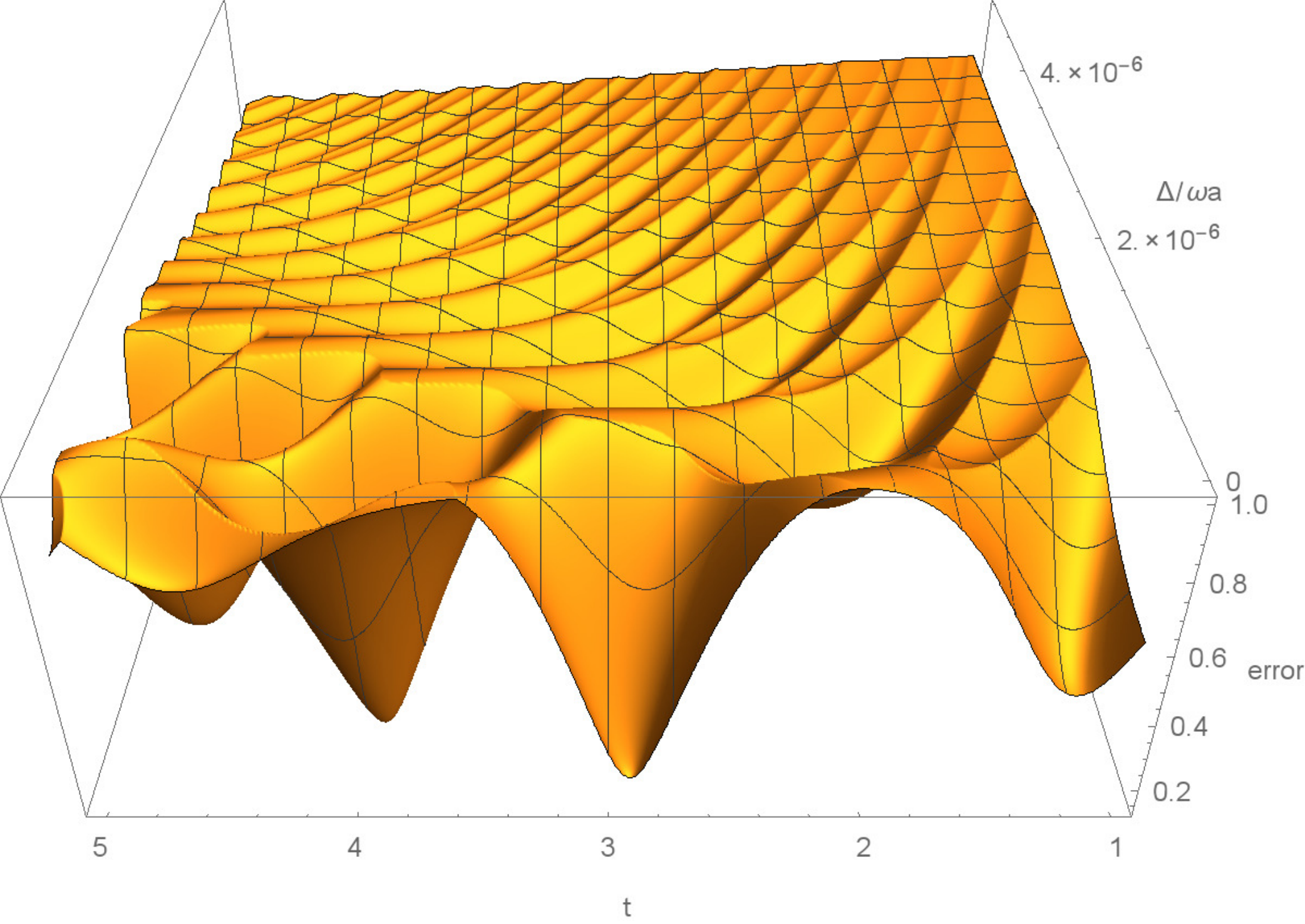}
\end{center}
 \caption{ \label{fig:detuning-3d} Error rate over detuning and time. }
\end{figure}

It was also shown that non-zero detuning levels could greatly improve the validity of the gate. On figure \ref{fig:detuning-3d}, it could be seen how does the error rate depend on the absolute detuning value. The graph is done only for low times for clarity, the picture is the same for higher times, but the  range of detuning values that interests us grows with time and also becomes more fine-grained.

It is visible that for values of $t$ around $4$, some non-zero detuning range actually lowers the overall error level. In the case of non-zero detuning, $t$ is not required to be a whole number, and it is not whole for most of the optimal parameters set.

\begin{figure}[H]
\begin{tabular}{c c}
\includegraphics[width=0.5\textwidth]{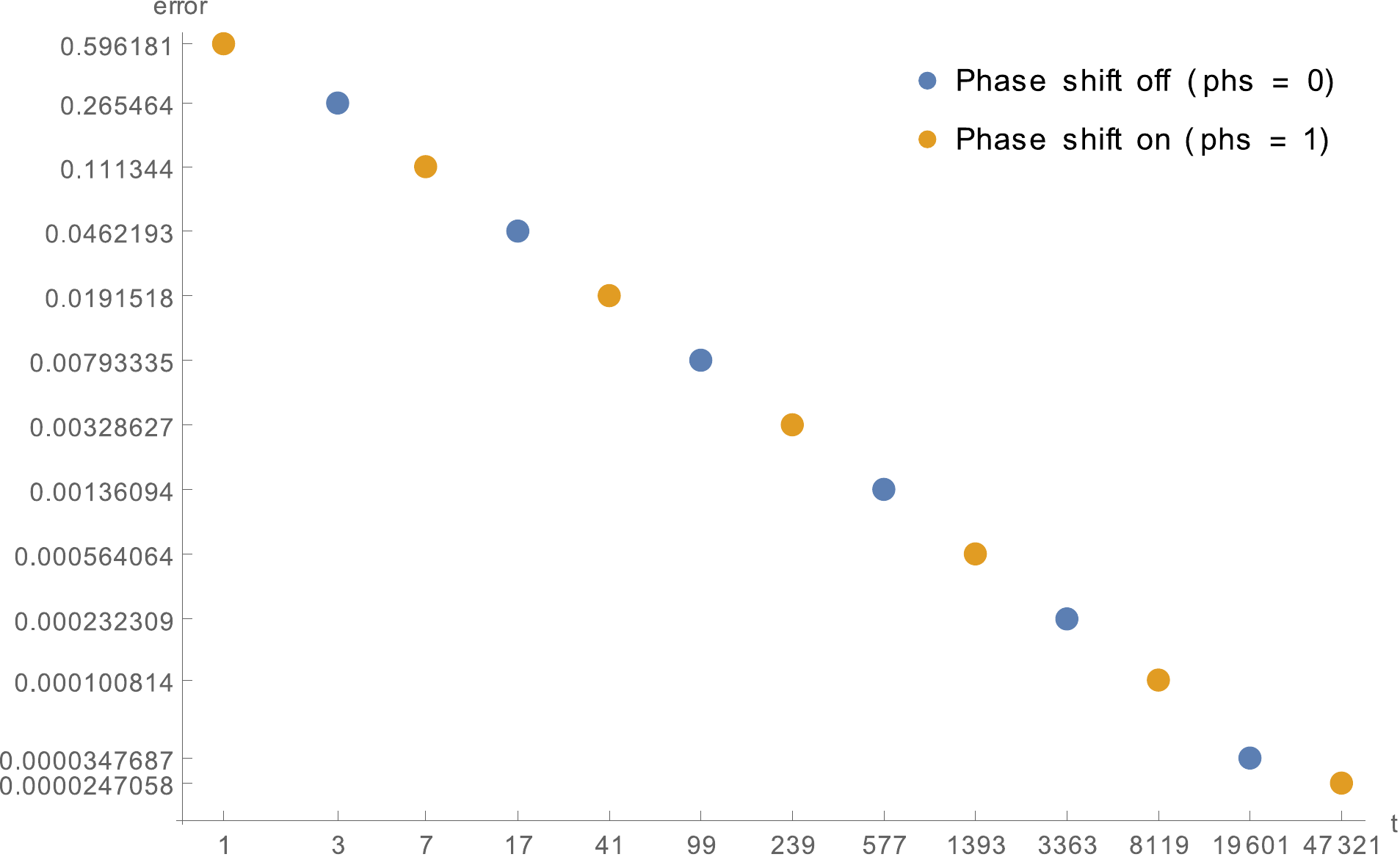} & \includegraphics[width=0.5\textwidth]{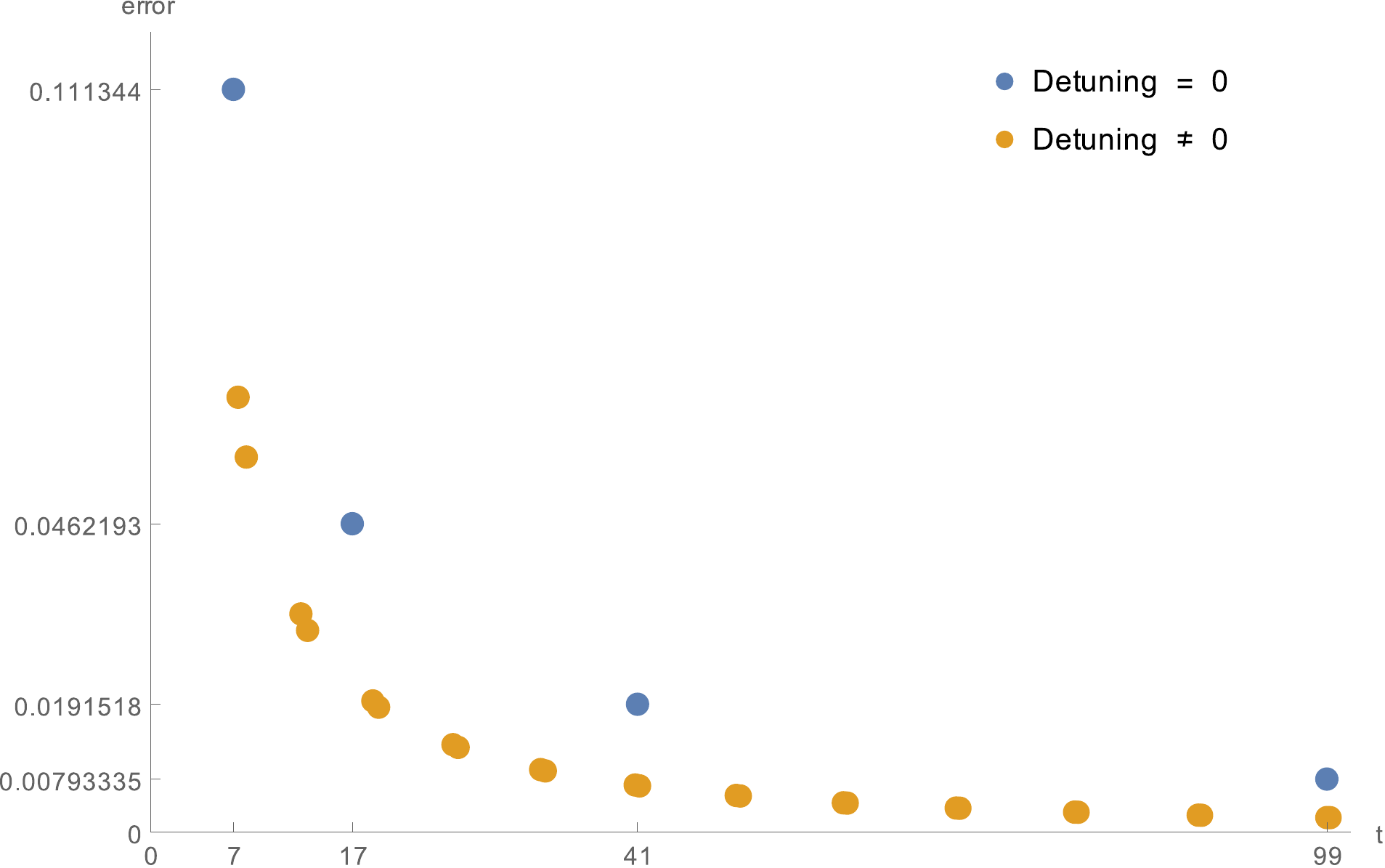} \\
Optimal cases without desired detuning & Optimal cases with desired detuning \\
(logarithmic scale on both axes) & (linear scale) \\
\end{tabular}
 \caption{ \label{fig:time-optimal} Optimal time values and their error rates for the cases with and without the desired detuning.  }
\end{figure}

If parameter sets with non-zero detuning value is taken into consideration, it greatly widens our possible selection of the ``optimal`` parameter sets, and also improves the lowest reachable error rate for a given time range. This could be seen on figure \ref{fig:time-optimal}. Also the same figure shows the overall dependency of the reachable error rate over our duration range.

\begin{figure}[H]
\begin{tabular}{c c}
\includegraphics[width=0.55\textwidth]{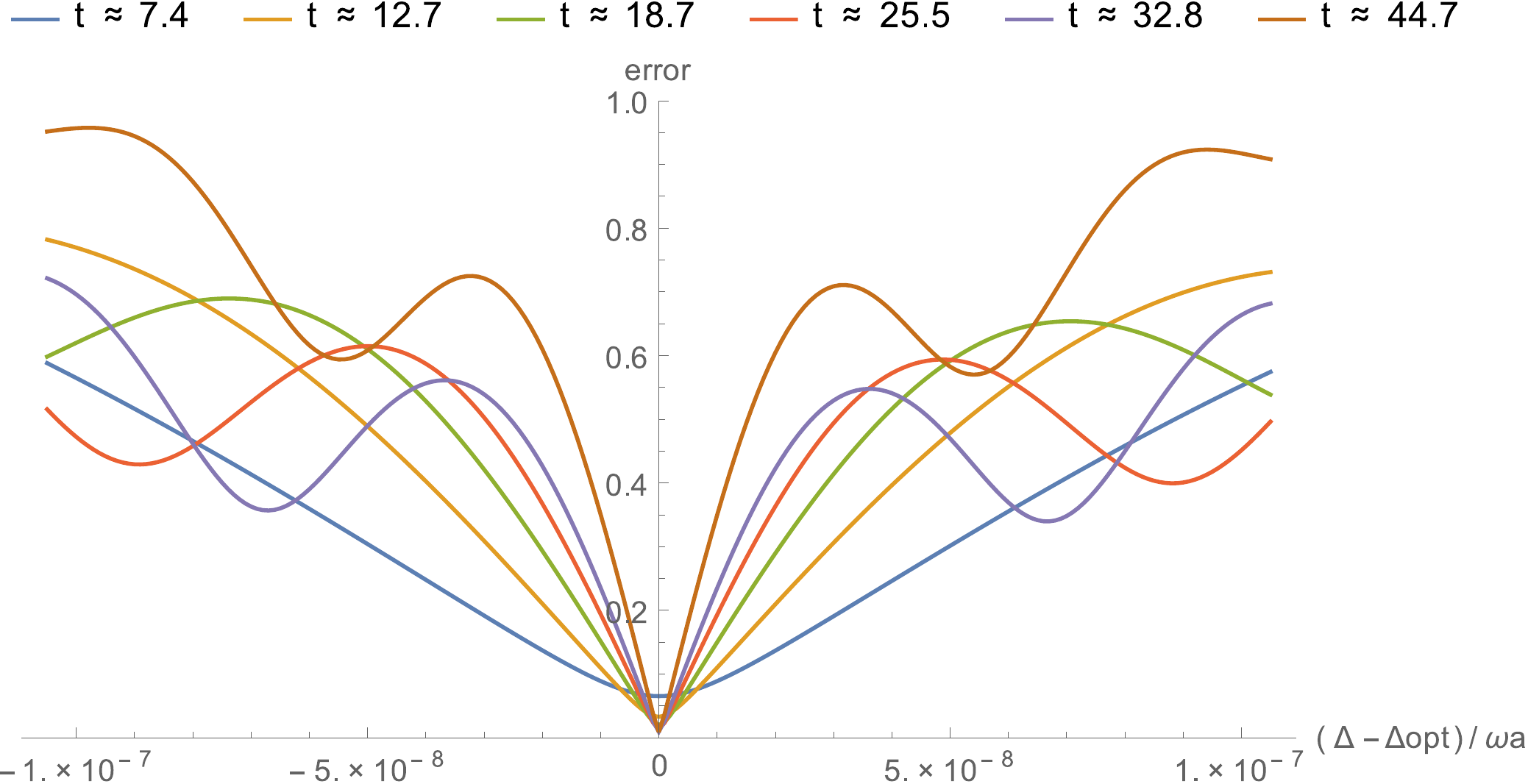} &
\includegraphics[width=0.45\textwidth]{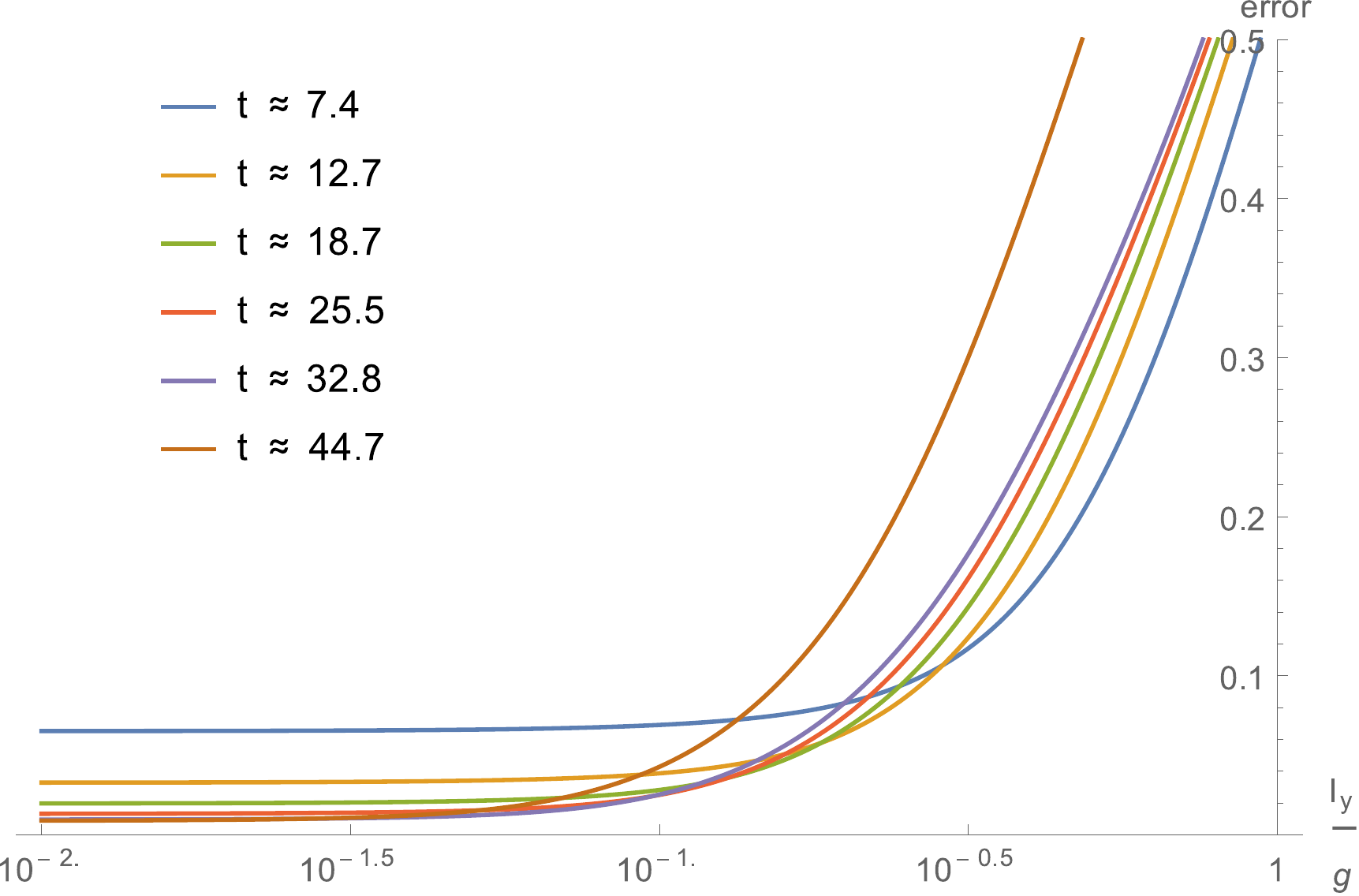} \\
Error over detuning offset & Error over photon leakage \\
& (logarithmic scale on leakage) \\
\end{tabular}
 \caption{ \label{fig:detuning-1} Error rates over detuning and photon leakage for the optimal cases with desired detuning. }
\end{figure}

Figure \ref{fig:detuning-1} shows how does the error rate in the cases of zero detuning depend on the inaccurate detuning setting $\Delta-\Delta_{opt}$ and from the photon emission coefficient. It could be seen that the dependency is not stronger than the corresponding dependency on figure \ref{fig:detuning-0}, so non-zero detuning does not introduce larger error values in cases of inaccurate detuning setting and/or non-zero photon leakage from the $NS$ gates cavities.

With $t = 99$, the lowest reachable error rate is $\approx 0.008$ without detuning and $\approx 0.002$ with $\Delta \approx 4.80 * g$.

\section{Conclusions}

We have conducted an estimation of the error rate of modified KLM model. Optimal time values were found for the zero detuning case, and it was shown how non-zero detuning could lower the error rate and broaden our choice for the time values.

Error introduced by the photon leakage from the cavity at the $NS$ gates was simulated, and it was shown how the photon emission rate affects the optimal time value.

Error introduced by inaccurate detuning setting was simulated, it was shown that it depends mostly on the chosen time.

Optimal $(time, detuning)$ pairs were found for the cases of non-zero detuning, and it was shown that those are more efficient (introduce a lower error for the same or lower duration) compared to the cases when the detuning is equal to zero.

It was also shown that the error introduced by photon leakage from the cavities and by the inaccurate detuning setting imprecision does not increase in the non-zero optimal detuning case when compared to zero detuning case. Thus, to get the optimal gate validity for the known photon escape rate and detuning setting imprecision, one could first calculate the maximum $NS$ gate duration after which the error introduced by them raises to an unacceptable level, and then find the optimal values for $t$ and $\frac{\Delta}{g}$ where $t$ is slightly less than that critical value.

The choice of the duration $\tau_0$ can be made by exact fixation of the moment $t_0$ when atom entries the cavity; for the speed of atoms about $1 cm.\ per\ sec.$ taken from a reservoir of cold gas such fixation can be performed with high precision.

Graphs at the figures  \ref{fig:time-optimal}, \ref{fig:detuning-1} shows that the lowest relative error level accessible by the considered model of C-Sign is about $0.01$; for the increasing noise it rapidly increases. If we roughly estimate for what number of qubits Grover search algorithm (\cite{G}) can be realized on this model, taking into account the linear growth of error in course of quantum computation and take the error on one Grover operator equal to C-Sign error we found, we obtain the number $n\approx 13$. This estimation, coming for the account of different sources of decoherence is more pessimistic than the more abstract estimations (see, for example, \cite{BC}) that gives the values $n\approx 20$. To reach such values in the realistic models, like KLM, we need the following additional modernization of this scheme.

\newpage

\section{Acknowledgements}

The work was partially supported by Russian Foundation for Basic Researches, grant 15-01-06132 a.


\begin{thebibliography}{99}

\bibitem{SH}P.W. Shor, Polynomial-Time Algorithms for Prime Factorization and Discrete Logarithms on a Quantum Computer, SIAM J. Comput. 26 (5): 1484–1509(1997).
\bibitem{G}L.K. Grover, A fast quantum mechanical algorithm for database search, Proceedings, 28th Annual ACM Symposium on the Theory of Computing, p. 212 (May 1996).
\bibitem{KLM}E. Knill, R. Laflamme, G. J. Milburn, A scheme for efficient quantum computation with linear optics, Nature 409, 46-52 (4 January 2001) | doi:10.1038/35051009.
\bibitem{KM}P. Kok, W.J. Munro, K. Nemoto, T.C. Ralph, Jonathan P. Dowling, G.J. Milburn, Review article: Linear optical quantum computing, Rev. Mod. Phys. 79, 135 (2007).
\bibitem{ML}C.R. Myers, R. Laflamme, Linear Optics Quantum Computation: an Overview, http://xxx.lanl.gov/abs/quant-ph/0512104 .
\bibitem{MP} A. Migdall, S. Polyakov, F. Jingyun, J. Bienfang, Single-Photon Generation and Detection: Physics and Applications, Volume 45 Experimental methods in the physical Sciences, Editorial Elsevier Inc. (2013).
\bibitem{H}S. Haroche, J. Raimond, Exploring the Quantum: Atoms, Cavities, and Photons. Oxford University Press (2006).
\bibitem{GC}D. Gottesman, Isaac L. Chuang, Quantum Teleportation is a Universal Computational Primitive, Nature 402, 390-393 (1999).
\bibitem{R} T. Ralph, Quantum optical systems for the implementation information processing, Progress in Physics, pp 69 (2006).
\bibitem{NC}M. Nielsen, I. Chuang, Quantum Computation and Quantum Information (10th Anniversary ed.), Cambridge University Press (2010).
\bibitem{RK}A. Reiserer, N. Kalb, G. Rempe, S. Ritter, A quantum gate between a flying optical photon and a single trapped atom, Nature 508, 237–240 (10 April 2014)
\bibitem{T}Q. A. Turchette, C. J. Hood, W. Lange, H. Mabuchi, and H. J. Kimble, Measurement of Conditional Phase Shifts for Quantum Logic, Phys. Rev. Lett. 75, 4710 
\bibitem{A}H. Azuma, Quantum computation with the Jaynes-Cummings model, Prog. Theor. Phys. 126 (2011), 369-385.
\bibitem{BC}Yu.I. Bogdanov, A.Yu. Chernyavskiy, B.I. Bantysh, V.F. Lukichev, A.A. Orlikovsky, I.A. Semenihin, D.V.Fastovets, A.S. Holevo, Numerical and analytical research of the impact of decoherence on quantum circuits, report for the International Symposium "Quantum Informatics-2014" (QI-2014), Zvenigorod, Moscow region, October 06-10, 2014, Proceedings of SPIE, 2014, vol. 9440.
\bibitem{JC}E.T. Jaynes, F.W. Cummings, Comparison of quantum and semiclassical radiation theories with application to the beam maser, Proc. IEEE 51 (1): 89–109, (1963). doi:10.1109/PROC.1963.1664

 



\end{thebibliography}
\end{document}